\documentclass[5pt]{article}
\usepackage{amsmath}
\usepackage[latin1]{inputenc}
\usepackage{amsmath,amsthm,amssymb}
\usepackage{graphicx}
\usepackage{color}

\renewcommand{\theequation}{\thesection.\arabic{equation}}

\newtheorem{lemma}{Lemma}[section]
\newtheorem{thm}{Theorem} [section]

\newtheorem{exmp}{Example} [section]
\newtheorem{cor}{Corollary}[section]
\newtheorem{rem}{Remark}[section]
\title{Two Constructions for Minimal Ternary Linear Codes
\thanks {The work of H.B. Liu is supported by the Natural Science Foundation of China with No.11901062, the work of Q.Y. Liao is supported by the Natural Science Foundation of China with No.12071321.}}
\author{{Haibo Liu\footnote{\ H.B.Liu, School of Applied Mathematics, Chengdu University of Information Technology,
 \small Chengdu, Sichuan, China (Email:liuhaibo@cuit.edu.cn)}
\quad Qunying Liao\footnote{\ Q.Y.Liao, School of Mathematical Sciences, Sichuan Normal University,
 \small Chengdu, Sichuan, China (Email:qunyingliao@sicnu.edu.cn)}
\quad Canze Zhu\footnote{\ C.Z.Zhu, School of Mathematical Sciences, Sichuan Normal University,
 \small Chengdu, Sichuan, China (Email:canzezhu@163.com)}}\ }
\date{}
\begin{document}
\baselineskip15pt \maketitle
\renewcommand{\theequation}{\arabic{section}.\arabic{equation}}
\catcode`@=11 \@addtoreset{equation}{section} \catcode`@=12
\begin{abstract}

Recently, minimal linear codes have been extensively studied due to their applications in secret sharing schemes, two-party computations, and so on. Constructing minimal linear codes violating the Ashikhmin-Barg condition and then determining their weight distributions have been interesting in coding theory and cryptography. In this paper, basing on exponential sums, Krawtchouk polynomials, and a function defined on special sets of vectors in $\mathbb{F}_3^m$, we present two new classes of minimal ternary linear codes violating the Ashikhmin-Barg condition, and then determine their complete weight enumerators. Especially, the minimal distance of a class of these codes is better than that of codes constructed in \cite{Heng-Ding-Zhou}.


 \end{abstract}

 \section{Introduction}
 $ $

Throughout the whole paper, let $p$ be a prime and $q=p^m$ for some positive integer $m$, denote $\mathbb{F}_p$ to be the finite field with $p$ elements. An $[n,k,d]$ linear code $\mathcal{C}$ over $\mathbb{F}_ p$ is a $k$-dimensional subspace of $\mathbb{F}_p^n$ with minimum (Hamming) distance $d$. For any $i=1,\dots,n$, $A_i$ denotes the number of codewords with Hamming weight $i$ in $\mathcal{C}$ of length $n$. The $\emph{weight}$ $\emph{enummerator}$ of $\mathcal{C}$ is defined by
\[1+A_1z+A_2z^2+\cdots+A_nz^n.\]
The $\emph{weight}$ $\emph{distribution}$ $(1,A_1,\dots,A_n)$ is an important research topic in coding theory, since it contains some crucial information as to estimate the error-correcting capability and the probability of error-detection and correction with respect to some algorithms \cite{Ding-Ding}. $\mathcal{C}$ is said to be a $t$-weight code if the number of nonzero $A_i$ in the sequence $(A_1,\dots,A_n)$ is equal to $t$. For a codeword $\mathbf{c}=(c_1,\dots,c_n)\in \mathcal{C}$, the complete weight enumerator of $\mathbf{c}$ is the monomial
\[w(\mathbf{c})=w_0^{t_0}w_1^{t_1}\cdots w_{p-1}^{t_{p-1}}\]
in the variables $w_0,w_1,\dots,w_{p-1}$, where $t_i(0\leq i\leq p-1)$ is the number of components of $\mathbf{c}$ equal to $w_i$. Then the complete weight enumerator of  $\mathcal{C}$ is \[CWE(\mathcal{C})=\sum_{\mathbf{c}\in \mathcal{C}}w(\mathbf{c}).\]

For a vector $\mathbf{a}=(a_1,\dots,a_n)\in \mathbb{F}_p^n$, the $\emph{S}upport$ of $\mathbf{a}$ is defined by
\[\text{Supp}(\mathbf{a})=\{1\leq i\leq n:a_i\neq 0\}.\]
Let $wt(\mathbf{a})$ be the Hamming weight of $\mathbf{a}$, then $wt(\mathbf{a})=|\text{Supp}(\mathbf{a})|$. For $\mathbf{b}\in \mathbb{F}_3^m$, we say that $\mathbf{a}$ covers $\mathbf{b}$ if $\text{Supp}(\mathbf{b})\subseteq \text{Supp}(\mathbf{a})$. A codeword $\mathbf{c}$ in $\mathcal{C}$ is minimal if $\mathbf{c}$ covers only those codewords $u\mathbf{c}$ $(u\in \mathbb{F}_p^\ast)$. $C$ is said to be minimal if every codeword in $\mathcal{C}$ is minimal.

It's well-known that minimal linear codes are wildly applied, especially in secret sharing schemes and two-party computations \cite{Carlet-Ding-Yuan}\cite{Yuan-Ding}. A sufficient condition for a linear code to be minimal is given in the following lemma \cite{Ashikhmin-Barg}.
\begin{lemma}(Ashikhmin-Barg)\label{lem9}
A linear code $\mathcal{C}$ over $\mathbb{F}_p$ is minimal if
\[\frac{w_{min}}{w_{max}}>\frac{p-1}{p},\]
where $w_{min}$ and $w_{max}$ denote the minimum and maximum nonzero Hamming weights for $\mathcal{C}$, respectively.
\end{lemma}
With the help of Lemma \ref{lem9}, many minimal linear codes were constructed from linear codes with a few weights \cite{Ding1} \cite{Ding2} \cite{Ding-Ding} \cite{Tang-Li-Qi-Zhou-Helleseth} \cite{Xiang}. The sufficient condition in Lemma \ref{lem9} is not usually necessary for a linear code to be minimal. Recently, searching for minimal linear codes with $\frac{w_{min}}{w_{max}}\leq \frac{p-1}{p}$ has been an interesting research topic. In 2018, Chang and Hyun \cite{Chang-Hyun} made a breakthrough and constructed an infinite family of minimal binary linear codes with $\frac{w_{min}}{w_{max}}< \frac{1}{2}$ by the generic construction
\begin{eqnarray}\label{shizi6}
\mathcal{C}_f=\{(uf(\mathbf{x})+\mathbf{v}\cdot \mathbf{x})_{\mathbf{x}\in {\mathbb{F}_p^m}^\ast}:u\in \mathbb{F}_p,\mathbf{v}\in\mathbb{F}_p^m\}.
\end{eqnarray}

 Based on the generic construction, a lot of minimal linear codes are obtained, which are not satisfying the Ashikhmin-Barg condition. Ding et al. \cite{Ding-Heng-Zhou} gave a necessary and sufficient condition for a binary linear code to be minimal, and employed special Boolean functions to obtain three classes of minimal binary linear codes. Heng et al. \cite{Heng-Ding-Zhou} used the characteristic function of a subset in $\mathbb{F}_3^m$ to construct a class of minimal ternary linear codes. Bartoli and Bonini \cite{Bartoli-Bonini} generalized the construction of minimal linear codes in \cite{Heng-Ding-Zhou} from ternary case to be odd characteristic case. Bonini and Borello \cite{Bonini-Borello} presented a family of minimal codes arising from cutting blocking sets. Tao et al. \cite{Tao-Feng-Li} obtained three-weight or four-weight minimal linear codes by using partial difference sets. Meanwhile, there are other methods to construct minimal linear codes violating the Ashikhmin-Barg condition, see  \cite{Bartoli-Bonini-Gunes} \cite{Li-Yue} \cite{Lu-Wu-Cao} \cite{Tang-Qi-Liao-Zhou} \cite{Xu-Qu}.

Till now, a lot of minimal linear codes violating the Ashikhmin-Barg condition are constructed, and their weight distributions are given, but their complete weight enumerator cannot be determined. In this paper, basing on the generic construction (\ref{shizi6}), and a special function defined on special sets of vectors in $\mathbb{F}_3^m$, we present two new classes of minimal ternary linear codes violating the Ashikhmin-Barg condition and then determine their complete weight enumerators. Especially, we show that the minimal distance of a class of these codes is better than that of codes constructed in \cite{Heng-Ding-Zhou}.

The paper is organized as follows. Section 2 provides some properties for Krawtchouk ploynomials and some results about minimal ternary codes, which will be needed in the sequel. Section 3 presents two classes of minimal ternary codes violating the Ashikhmin-Barg condition, and then determines their complete weight enumerators. Section 4 concludes the whole paper.

\section{Preliminaries}
$ $

the Krawtchouk polynomial introduced by Lloyd in 1957 \cite{Lloyd} is applied in coding theory \cite{Best} \cite{Fu-Klove-Luo-Wei}, cryptography \cite{Carlet-Zeng-Li-Hu} and combinatorics \cite{Levenshtein}. Here we only give a brief introduction to the Krawtchouk polynomial with the essential properties. For more details, readers are referred to \cite{Best} \cite{Fu-Klove-Luo-Wei} \cite{Levenshtein} \cite{Lloyd}.

Let $m$, $h$ be positive integers and $x$ be a variable taking nonnegative values. The \emph{Krawtchouk polynomial }(of degree $t$ with parameters $h$ and $m$ ) is defined by
\[K_t(x,m)=\sum_{j=0}^t(-1)^j(h-1)^{t-j}{x\choose j}{{m-x}\choose {t-j}}.\]
Accordingly, the \emph{Lloyd polynomial } $\Psi_k(x,m)$ (of degree $t$ with parameters $h$ and $m$) is given by
\[\Psi_k(x,m)=\sum_{t=0}^kK_t(x,m).\]

The following Lemma \ref{lem4} will be useful in the sequel.
\begin{lemma}\cite{Heng-Ding-Zhou}\label{lem4}
Let symbols and notations be as the above, suppose that $1\leq x\leq m$, $1\leq k\leq m-1$, $\mathbf{u}\in \mathbb{Z}_q^m$ with Hamming weight $wt(\mathbf{u})=i$, then the following hold:

(1) $\Psi_k(x,m)=K_k(x-1,m-1)$;

(2) $K_t(0,m)=(h-1)^t{m\choose t}$;

(3) $|\Psi_k(x,m)|\leq(h-1)^k{{m-1}\choose k}$;

(4) $\sum_{\mathbf{v}\in  \mathbb{Z}_q^m,wt(\mathbf{v})=i}\zeta_q^{\mathbf{u}\cdot\mathbf{v}}=K_t(i,m)$,\\
where $\zeta_q$ denotes the $q$-th primitive root of complex unity, and the inner product $\mathbf{u}\cdot\mathbf{v}$ in $\mathbb{Z}_q^m$  is defined by $$\mathbf{u}\cdot\mathbf{v}=u_1v_1+\cdots+u_mv_m.$$
\end{lemma}
Note that the upper bound for $|\Psi_k(x,m)|$ in Lemma \ref{lem4} is tight since
\[\Psi_k(1,m)=K_k(0,m-1)=(h-1)^k{{m-1}\choose k}.\]

Assume $f(\mathbf{x})$ is a function from $\mathbb{F}_p^m$ to $\mathbb{F}_p$ such that $f(\mathbf{0})=0$ and $f(\mathbf{b})\neq 0$ for at least one $\mathbf{b}\in \mathbb{F}_p^m$. Recall that the Walsh transform of $f$ is given by
\[\widehat{f}(\mathbf{w})=\sum_{\mathbf{x}\in\mathbb{F}_p^m}\zeta_p^{f(\mathbf{x})-\mathbf{w}\cdot \mathbf{x}}\quad (\mathbf{w}\in \mathbb{F}_p^m).\]

The following result shows that for $p=3$ the weight distribution of the ternary code $\mathcal{C}_f$ can be determined by the Walsh spectrum of $f$.
\begin{lemma}\cite{Heng-Ding-Zhou}\label{lem5}
For $p=3$, assume that $f(\mathbf{x})\neq \mathbf{w}\cdot \mathbf{x}$ for any $\mathbf{w}\in \mathbb{F}_3^m$. Then the linear code $\mathcal{C}_f$ in (\ref{shizi6}) has length $3^m-1$ and dimension $m+1$. In addition, the weight distribution of $\mathcal{C}_f$ is given by the following multiset union:
\[\{\{2(3^{m-1}-\frac{Re(\widehat{f}(\mathbf{v}))}{3}):u\in \mathbb{F}_3^\ast,\mathbf{v}\in \mathbb{F}_3^m\}\}\bigcup\]
\[\{\{3^m-3^{m-1}:u=0,\mathbf{v}\in{\mathbb{F}_3^m}^\ast\}\}\bigcup\{\{0\}\}.\]
Herein and hereafter, $Re(e)$ denotes the real part of the complex number $e$.
\end{lemma}

The following Lemma \ref{lem8} gives a sufficient and necessary condition for $\mathcal{C}_f$ to be minimal in terms of the Walsh spectrum of $f$.
\begin{lemma}\cite{Heng-Ding-Zhou}\label{lem8}
Let $\mathcal{C}_f$ be the ternary code in Lemma \ref{lem5}. Assume that $f(\mathbf{x})\neq \mathbf{v}\cdot \mathbf{x}$ for any $\mathbf{v}\in \mathbb{F}_3^m$. Then $\mathcal{C}_f$ is a $[3^m-1,m+1]$ minimal code if and only if both
\[Re(\widehat{f}(\mathbf{w}_1))+Re(\widehat{f}(\mathbf{w}_2))-2Re(\widehat{f}(\mathbf{w}_3))\neq 3^m\]
and
\[Re(\widehat{f}(\mathbf{w}_1))+Re(\widehat{f}(\mathbf{w}_2))+Re(\widehat{f}(\mathbf{w}_3))\neq 3^m\]
for any pairwise distinct vectors $\mathbf{w}_1$, $\mathbf{w}_2$ and $\mathbf{w}_3$ in $\mathbb{F}_3^m$ satisfying $\mathbf{w}_1+\mathbf{w}_2+\mathbf{w}_3=\mathbf{0}$.
\end{lemma}
For a positive integer $k$ with $1\leq k\leq m$, let $S(m,k)$ denote the set of vectors in $\mathbb{F}_3^m \setminus\{0\}$ with Hamming weight at most $k$. It's clear that
\[|S(m,k)|=\sum_{j=1}^k2^j{m\choose j}.\]
Define the function $g_{(m,k)}:\mathbb{F}_3^m\rightarrow \mathbb{F}_3$ as
\begin{eqnarray*}
g_{(m,k)}(\mathbf{x})=\left\{\begin{array}{ll}1, &\text{if}\quad \mathbf{x}\in S(m,k);\\
 0, &\text{otherwise}. \end{array}\right.
\end{eqnarray*}
 Taking $f=g_{(m,k)}$ in (\ref{shizi6}), one can obtain a ternary linear code $C_{g_{(m,k)}}$, which is a minimal linear code violating the Ashikhmin-Barg condition. The parameters and the weight distribution of $C_{g_{(m,k)}}$ are given as follows, which will be needed in next section.
\begin{lemma}\cite{Heng-Ding-Zhou}\label{lem3}
Let $m,k$ be integers with $m\geq 5$ and $2\leq k\leq \lfloor\frac{m-1}{2}\rfloor$. Then the linear code $C_{g_{(m,k)}}$ is minimal with parameters
\[\left[3^m-1,m+1,\sum_{j=1}^k2^j{m\choose j}\right],\]
and the weight distribution is given by Table I, where $\Psi_k(i,m)$ is the Lloyd polynomial. Furthermore, $\frac{w_{min}}{w_{max}}\leq \frac{2}{3}$ if and only if
\[3\sum_{j=1}^k2^j{m\choose j}\leq 2(3^m-3^{m-1})+2^{k+1}{{m-1}\choose k}-2.\]
\end{lemma}

\[\text{\small Table I: the weight distribution of $C_{g_{(m,k)}}$ in Lemma \ref{lem3}}\]
$$
\scalebox{0.9}{

\begin{tabular}{|l|l|l|}\hline
Weight $w$ & Multiplicity $A_{w}$ &condition  \\\hline
0 &1 &$u=0,\mathbf{v}=\mathbf{0}$\\\cline{1-3}
$3^m-3^{m-1}+\Psi_k(i,m)-1$ &$2^{1+i}{m\choose i}(1\leq i\leq m)$&$u\in\mathbb{F}_3^\ast,wt(\mathbf{v})=i$ with $Re(\widehat{g}_{(m,k)}(\mathbf{v}))=-\frac{3}{2}(\Psi_k(i,m)-1)$\\\cline{1-3}
 $\sum_{j=1}^k2^j{m\choose j}$&$2$&$u\in\mathbb{F}_3^\ast,\mathbf{v}=\mathbf{0}$ with $Re(\widehat{g}_{(m,k)}(\mathbf{0}))=3^m-\frac{3}{2}\sum_{j=1}^k2^j{m\choose j}$\\\cline{1-3}
  $3^m-3^{m-1}$&$3^m-1$&$u=0,\mathbf{v}\in{\mathbb{F}_3^m}^\ast$\\\hline
\end{tabular}}\\
$$

Let $D\subseteq \mathbb{F}_q^\ast$ and $\overline{D}=\mathbb{F}_q^\ast \setminus D$, then the characteristic function of $D$ is defined to be
\begin{eqnarray*}
f_D(\mathbf{x})=\left\{\begin{array}{ll}1, &\text{if}\quad \mathbf{x}\in D;\\
 0, &\text{otherwise}. \end{array}\right.
\end{eqnarray*}

And the relationship between $\widehat{f}_D(\mathbf{w})$ and $\widehat{f}_{\overline{D}}(\mathbf{w})$ is given by the following lemma.
\begin{lemma}\cite{Mesnager}\label{lem6}
\begin{eqnarray*}
\widehat{f}_D(\mathbf{w})+\widehat{f}_{\overline{D}}(\mathbf{w})=\left\{\begin{array}{ll}(q-1)\zeta_p+q+1, &\text{if}\quad \mathbf{w}=0;\\
 1-\zeta_p, &\text{otherwise}. \end{array}\right.
\end{eqnarray*}
\end{lemma}

\section{Main Results and Their Proofs}
$ $

In this section, we first prove that the characteristic function $g_{(m,k)}$ in Lemma \ref{lem3} can be generalized to be $f_{(m,k)}$ for obtaining a class of minimal linear codes violating the Ashikhmin-Barg condition.

Let $S$ be a nonempty subset of $[k]=\{1,\dots,k\}$ and $A=\{\mathbf{x}\in \mathbb{F}_3^m| wt(\mathbf{x})\in S\}$, define the function $f_{(m,k)}$ from $\mathbb{F}_3^m$ to $\mathbb{F}_3$ as
\begin{eqnarray*}
f_{(m,k)}(\mathbf{x})=\left\{\begin{array}{lll}-1, &\text{if}\quad wt(\mathbf{x})\in S;\\
1, &\text{if}\quad wt(\mathbf{x})\in [k]\setminus  S;\\
 0, &\text{otherwise}. \end{array}\right.
\end{eqnarray*}
\begin{thm}\label{thm1}
Let $m,k$ be integers with $m\geq 5$ and $2\leq k\leq \lfloor\frac{m-1}{2}\rfloor$. Then the linear code $C_{f_{(m,k)}}$ is minimal with parameters
\[\left[3^m-1,m+1,\sum_{j=1}^k2^j{m\choose j}\right],\]
and the weight distribution is given by Table II, and the complete weight enumerator is
\begin{eqnarray*}
&&\omega_0^{3^{m}-1}+(3^m-1)\omega_0^{3^{m-1}-1}\omega_1^{3^{m-1}}\omega_2^{3^{m-1}}\\
&&\omega_0^{3^{m-1}-\sum_{i=1}^k2^i-1}\omega_1^{\sum_{i=1}^m2^i-|A|}\omega_2^{|A|}+\omega_0^{3^{m-1}-\sum_{i=1}^k2^i-1}\omega_1^{|A|}
\omega_2^{\sum_{i=1}^m2^i-|A|}\\
&&\sum_{i=1}^m2^i{m\choose i}\omega_0^{3^{m-1}-\Psi_k(i,m)}\omega_1^{3^{m-1}+\Psi_k(i,m)-\sum_{t\in S}K_t(i,m)-1}\omega_2^{3^{m-1}+\sum_{t\in S}K_t(i,m)}\\
&&\sum_{i=1}^m2^i{m\choose i}\omega_0^{3^{m-1}-\Psi_k(i,m)}\omega_1^{3^{m-1}+\sum_{t\in S}K_t(i,m)}\omega_2^{3^{m-1}+\Psi_k(i,m)-\sum_{t\in S}K_t(i,m)-1},
\end{eqnarray*}
where $\Psi_k(i,m)$ is the Lloyd polynomial. Furthermore, $\frac{w_{min}}{w_{max}}\leq \frac{2}{3}$ if and only if
\[3\sum_{j=1}^k2^j{m\choose j}\leq 2(3^m-3^{m-1})+2^{k+1}{{m-1}\choose k}-2.\]
\end{thm}
\[\text{\small Table II: the weight distribution of $C_{f_{(m,k)}}$ in Theorem \ref{thm1}}\]
$$
\scalebox{0.9}{

\begin{tabular}{|l|l|l|}\hline
Weight $w$ & Multiplicity $A_{w}$ &condition  \\\hline
0 &1 &$u=0,\mathbf{v}=\mathbf{0}$\\\cline{1-3}
$3^m-3^{m-1}+\Psi_k(i,m)-1$ &$2^{1+i}{m\choose i}(1\leq i\leq m)$&$u\in\mathbb{F}_3^\ast,wt(\mathbf{v})=i$ with $Re(\widehat{f}_{(m,k)}(\mathbf{v}))=-\frac{3}{2}(\Psi_k(i,m)-1)$\\\cline{1-3}
 $\sum_{j=1}^k2^j{m\choose j}$&$2$&$u\in\mathbb{F}_3^\ast,\mathbf{v}=\mathbf{0}$ with $Re(\widehat{f}_{(m,k)}(\mathbf{0}))=3^m-\frac{3}{2}\sum_{j=1}^k2^j{m\choose j}$\\\cline{1-3}
  $3^m-3^{m-1}$&$3^m-1$&$u=0,\mathbf{v}\in{\mathbb{F}_3^m}^\ast$\\\hline
\end{tabular}}\\
$$

$\mathbf{Proof.}$ For any $\mathbf{v}\in \mathbb{F}_3^m$, to prove Theorem \ref{thm1}, it's enough to show that $Re(\widehat{f}_{(m,k)}(\mathbf{v}))$ equals to $Re(\widehat{g}_{(m,k)}(\mathbf{v}))$. In fact, from the definition of $f_{(m,k)}$, we have
\begin{eqnarray*}
\widehat{f}_{(m,k)}(\mathbf{v})&=&\sum_{\mathbf{x}\in\mathbb{F}_3^m}\zeta_3^{f_{(m,k)}(\mathbf{x})-\mathbf{v}\cdot \mathbf{x}}=\sum_{wt(\mathbf{x})\in S}\zeta_3^{-1-\mathbf{v}\cdot \mathbf{x}}+\sum_{wt(\mathbf{x})\in [k]\setminus  S}\zeta_3^{1-\mathbf{v}\cdot \mathbf{x}}+\sum_{\mathbf{x}\in \mathbb{F}_3^m \setminus S(m,k)}\zeta_3^{-\mathbf{v}\cdot \mathbf{x}}\\
&=&\sum_{\mathbf{x}\in\mathbb{F}_3^m}\zeta_3^{-\mathbf{v}\cdot \mathbf{x}}+\sum_{wt(\mathbf{x})\in S}(\zeta_3^{-1-\mathbf{v}\cdot \mathbf{x}}-\zeta_3^{-\mathbf{v}\cdot \mathbf{x}})+\sum_{wt(\mathbf{x})\in [k]\setminus  S}(\zeta_3^{1-\mathbf{v}\cdot \mathbf{x}}-\zeta_3^{-\mathbf{v}\cdot \mathbf{x}})\\
&=&\sum_{\mathbf{x}\in\mathbb{F}_3^m}\zeta_3^{-\mathbf{v}\cdot \mathbf{x}}+\sum_{wt(\mathbf{x})\in S }(\zeta_3^{-1}-1)\zeta_3^{-\mathbf{v}\cdot \mathbf{x}}+\sum_{wt(\mathbf{x})\in [k]\setminus  S}(\zeta_3-1)\zeta_3^{-\mathbf{v}\cdot \mathbf{x}}\\
&=&\sum_{\mathbf{x}\in\mathbb{F}_3^m}\zeta_3^{-\mathbf{v}\cdot \mathbf{x}}+\sum_{wt(\mathbf{x})\in [k]}(\zeta_3-1)\zeta_3^{-\mathbf{v}\cdot \mathbf{x}}+\sum_{wt(\mathbf{x})\in S}(\zeta_3^{-1}-\zeta_3)\zeta_3^{-\mathbf{v}\cdot \mathbf{x}}.
\end{eqnarray*}
If $\mathbf{v}=\mathbf{0}$, then we have
\[\widehat{f}_{(m,k)}(\mathbf{0})=3^m+(\zeta_3-1)\sum_{j=1}^k2^j{m\choose j}+(\zeta_3^{-1}-\zeta_3)|A|\]
and
\[Re(\widehat{f}_{(m,k)}(\mathbf{0}))=3^m-\frac{3}{2}\sum_{j=1}^k2^j{m\choose j}.\]
If $wt(\mathbf{v})=i(1\leq i\leq m)$, then by Lemma \ref{lem4} we have
\[\widehat{f}_{(m,k)}(\mathbf{v})=(\zeta_3-1)\sum_{t=1}^kK_t(i,m)+(\zeta_3^{-1}-\zeta_3)\sum_{t\in S}K_t(i,m)=(\zeta_3-1)(\Psi_k(i,m)-1)+(\zeta_3^{-1}-\zeta_3)\sum_{t\in S}K_t(i,m).\]
Thus
\[Re(\widehat{f}_{(m,k)}(\mathbf{v}))=-\frac{3}{2}((\Psi_k(i,m)-1)).\]
Then the weight distribution of $C_{f_{(m,k)}}$ follows from Lemma \ref{lem5}, and for any $\mathbf{v}\in \mathbb{F}_3^m$, $Re(\widehat{f}_{(m,k)}(\mathbf{v}))=Re(\widehat{g}_{(m,k)}(\mathbf{v}))$, thus the desired results follow from Lemma \ref{lem3}.

Next, we determine the complete weight enumerator of $C_{f_{(m,k)}}$. For any $\lambda, u\in\mathbb{F}_3$, and $\mathbf{v}\in \mathbb{F}_3^m$, denote
\[N_{\lambda(u,\mathbf{v})}=\{\mathbf{x}\in\mathbb{F}_3^m|uf_{(m,k)}(\mathbf{x})+\mathbf{v}\cdot \mathbf{x}=\lambda \},\]
then we have
\begin{eqnarray*}
|N_{\lambda(u,\mathbf{v})}|&=&\frac{1}{3}\sum_{z\in \mathbb{F}_3 }\sum_{\mathbf{x}\in\mathbb{F}_3^m}\zeta_3^{z(uf_{(m,k)}(\mathbf{x})+\mathbf{v}\cdot \mathbf{x}-\lambda)}\\
&=&3^{m-1}+\frac{1}{3}(\zeta_3^{-\lambda}\sum_{\mathbf{x}\in\mathbb{F}_3^m}\zeta_3^{uf_{(m,k)}(\mathbf{x})+\mathbf{v}\cdot \mathbf{x}}+\zeta_3^{\lambda}\sum_{\mathbf{x}\in\mathbb{F}_3^m}\zeta_3^{-uf_{(m,k)}(\mathbf{x})-\mathbf{v}\cdot \mathbf{x}})\\
&=&3^{m-1}+\frac{1}{3}\zeta_3^{-\lambda}(\sum_{wt(\mathbf{x})\in S}\zeta_3^{-u+\mathbf{v}\cdot \mathbf{x}}+\sum_{wt(\mathbf{x})\in [k]\setminus  S}\zeta_3^{u+\mathbf{v}\cdot \mathbf{x}}+\sum_{wt(\mathbf{x})\in\mathbb{F}_3^m\setminus [k]}\zeta_3^{\mathbf{v}\cdot \mathbf{x}})\\
&&+\frac{1}{3}\zeta_3^{\lambda}(\sum_{wt(\mathbf{x})\in S}\zeta_3^{u-\mathbf{v}\cdot \mathbf{x}}+\sum_{wt(\mathbf{x})\in [k]\setminus  S}\zeta_3^{-u-\mathbf{v}\cdot \mathbf{x}}+\sum_{wt(\mathbf{x})\in\mathbb{F}_3^m\setminus [k]}\zeta_3^{-\mathbf{v}\cdot \mathbf{x}})\\
&=&3^{m-1}+\frac{1}{3}\zeta_3^{-\lambda}(\sum_{wt(\mathbf{x})\in S}(\zeta_3^{-u}-\zeta_3^{u})\zeta_3^{\mathbf{v}\cdot \mathbf{x}}+\sum_{wt(\mathbf{x})\in [k]}
(\zeta_3^{u}-1)\zeta_3^{\mathbf{v}\cdot \mathbf{x}}+\sum_{\mathbf{x}\in\mathbb{F}_3^m}\zeta_3^{\mathbf{v}\cdot \mathbf{x}})\\
&&+\frac{1}{3}\zeta_3^{\lambda}(\sum_{wt(\mathbf{x})\in S}(\zeta_3^{u}-\zeta_3^{-u})\zeta_3^{\mathbf{v}\cdot \mathbf{x}}+\sum_{wt(\mathbf{x})\in [k]}(\zeta_3^{-u}-1)\zeta_3^{\mathbf{v}\cdot \mathbf{x}}+\sum_{\mathbf{x}\in\mathbb{F}_3^m}\zeta_3^{\mathbf{v}\cdot \mathbf{x}}).\qquad\qquad\qquad\qquad (1.2)\\
\end{eqnarray*}

Now we can calculate $|N_{\lambda(u,\mathbf{v})}|$ according to the following three cases.

{\bf Case 1}. If $u=0$, according to the above equation (1.2), we have
\begin{eqnarray*}
|N_{\lambda(0,\mathbf{v})}|&=&3^{m-1}+\frac{1}{3}(\zeta_3^{\lambda}+\zeta_3^{-\lambda})\sum_{\mathbf{x}\in\mathbb{F}_3^m}\zeta_3^{\mathbf{v}\cdot \mathbf{x}}
=\left\{\begin{array}{ll} 3^m, &\text{if}\quad \mathbf{v}=\mathbf{0};\\
   3^{m-1}, & \text{if}\quad \mathbf{v}\neq \mathbf{0}.
\end{array}\right.
\end{eqnarray*}

{\bf Case 2}. If $u\neq 0$ and $\mathbf{v}=\mathbf{0}$, then
\begin{eqnarray*}
|N_{\lambda(u,\mathbf{0})}|&=&3^{m-1}+\frac{1}{3}\zeta_3^{-\lambda}((\zeta_3^{-u}-\zeta_3^{u})|A|+
(\zeta_3^{u}-1)\sum_{i=1}^k2^{i}{m\choose i}+3^m)\\
&&+\frac{1}{3}\zeta_3^{\lambda}((\zeta_3^{u}-\zeta_3^{-u})|A|+(\zeta_3^{-u}-1)\sum_{i=1}^k2^{i}{m\choose i}+3^m)\\
&=&3^{m-1}+\frac{1}{3}((\zeta_3^{-\lambda}+\zeta_3^{\lambda})3^m+((\zeta_3^{-\lambda+u}+\zeta_3^{\lambda-u})-(\zeta_3^{-\lambda}+\zeta_3^{\lambda}))
\sum_{i=1}^k2^{i}{m\choose i}\\
&&+((\zeta_3^{-\lambda-u}+\zeta_3^{\lambda+u})-(\zeta_3^{-\lambda+u}+\zeta_3^{\lambda-u}))|A|)\\
&=&\left\{\begin{array}{lll}\sum_{i=1}^k2^{i}{m\choose i}-|A|, &\text{if}\quad u=\lambda;\\
|A|&\text{if}\quad u=-\lambda;\\
3^m-\sum_{i=1}^k2^{i}{m\choose i}, & \text{if}\quad \lambda=0.
\end{array}\right.
\end{eqnarray*}

{\bf Case 3}. If $u\neq 0$ and $wt(\mathbf{v})=i(1\leq i\leq m)$, then one can get
\begin{eqnarray*}
|N_{\lambda(u,\mathbf{v})}|&=&3^{m-1}+\frac{1}{3}\zeta_3^{-\lambda}(\sum_{wt(\mathbf{x})\in S}(\zeta_3^{-u}-\zeta_3^{u})\zeta_3^{\mathbf{v}\cdot \mathbf{x}}+\sum_{wt(\mathbf{x})\in [k]}(\zeta_3^{u}-1)\zeta_3^{\mathbf{v}\cdot \mathbf{x}})\\
&&+\frac{1}{3}\zeta_3^{\lambda}(\sum_{wt(\mathbf{x})\in S}(\zeta_3^{u}-\zeta_3^{-u})\zeta_3^{\mathbf{v}\cdot \mathbf{x}}+\sum_{wt(\mathbf{x})\in [k]}(\zeta_3^{-u}-1)\zeta_3^{\mathbf{v}\cdot \mathbf{x}})\\
&=&3^{m-1}+\frac{1}{3}((\zeta_3^{-\lambda+u}+\zeta_3^{\lambda-u})-(\zeta_3^{-\lambda}+\zeta_3^{\lambda}))\sum_{wt(\mathbf{x})\in [k]}\zeta_3^{\mathbf{v}\cdot \mathbf{x}}\\
&&+\frac{1}{3}((\zeta_3^{-\lambda-u}+\zeta_3^{\lambda+u})-(\zeta_3^{-\lambda+u}+\zeta_3^{\lambda-u}))\sum_{wt(\mathbf{x})\in S}\zeta_3^{\mathbf{v}\cdot \mathbf{x}}\\
&=&\left\{\begin{array}{lll}3^{m-1}+\Psi_k(i,m)-\sum_{t\in S}K_t(i,m)-1, &\text{if}\quad u=\lambda;\\
3^{m-1}+\sum_{t\in S}K_t(i,m), &\text{if}\quad u=-\lambda;\\
3^{m-1}-\Psi_k(i,m)+1, & \text{if}\quad \lambda=0.
\end{array}\right.
\end{eqnarray*}

By {\bf Cases 1-3}, note that $u\in \mathbb{F}_3$, thus we obtain the complete weight enumerator of $C_{f_{(m,k)}}$.

Summarizing the discussions above, we complete the proof for Theorem \ref{thm1}. \hfill$\Box$\\

\begin{rem}
In the similar calculation, the complete weight enumerator of $C_{g_{(m,k)}}$ in Lemma \ref{lem3} is
\begin{eqnarray*}
&&\omega_0^{3^{m}-1}+(3^m-1)\omega_0^{3^{m-1}-1}\omega_1^{3^{m-1}}\omega_2^{3^{m-1}}+
\omega_0^{3^{m-1}-\sum_{i=1}^k2^i-1}\omega_1^{\sum_{i=1}^m2^i}\omega_2^{0}+\omega_0^{3^{m-1}-\sum_{i=1}^k2^i-1}\omega_1^{0}\omega_2^{\sum_{i=1}^m2^i}\\
&&\sum_{i=1}^m2^i{m\choose i}\omega_0^{3^{m-1}-\Psi_k(i,m)}\omega_1^{3^{m-1}+\Psi_k(i,m)-1}\omega_2^{3^{m-1}}
+\sum_{i=1}^m2^i{m\choose i}\omega_0^{3^{m-1}-\Psi_k(i,m)}\omega_1^{3^{m-1}}\omega_2^{3^{m-1}+\Psi_k(i,m)-1},
\end{eqnarray*}
where $\Psi_k(i,m)$ is the Lloyd polynomial. This implies that the complete weight enumerator of $C_{f_{(m,k)}}$ is different from that of $C_{g_{(m,k)}}$, namely, $C_{f_{(m,k)}}$ and $C_{g_{(m,k)}}$ are different codes.
\end{rem}

Next, we construct a new class of minimal linear codes violating the Ashikhmin-Barg condition basing on Lemma \ref{lem5} and Lemma \ref{lem3}. Recall that $S(m,k)$ denotes the set of vectors in ${\mathbb{F}_3^m}^\ast$ with Hamming weight at most $k$, now set $\overline{S}(m,k)={\mathbb{F}_3^m}^\ast\setminus S(m,k)$, we can define the function $\overline{g}_{(m,k)}:\mathbb{F}_3^m\rightarrow \mathbb{F}_3$ as
\begin{eqnarray*}
\overline{g}_{(m,k)}(\mathbf{x})=\left\{\begin{array}{ll}1, &\text{if}\quad \mathbf{x}\in \overline{S}(m,k);\\
 0, &\text{otherwise}. \end{array}\right.
\end{eqnarray*}
\begin{thm}\label{thm2}
The ternary code $C_{\overline{g}_{(m,k)}}$ has length $3^m-1$ and dimension $m+1$, the weight distribution is given by Table III, and the complete weight enumerator is
\begin{eqnarray*}
&&\omega_0^{3^{m}-1}+(3^m-1)\omega_0^{3^{m-1}-1}\omega_1^{3^{m-1}}\omega_2^{3^{m-1}}+
\omega_0^{\sum_{i=0}^k2^i-1}\omega_1^{3^{m}-\sum_{i=0}^m2^i}\omega_2^{0}+\omega_0^{\sum_{i=0}^k2^i-1}\omega_1^{0}\omega_2^{3^m-\sum_{i=0}^m2^i}\\
&&\sum_{i=1}^m2^i{m\choose i}\omega_0^{3^{m-1}+\Psi_k(i,m)-1}\omega_1^{3^{m-1}-\Psi_k(i,m)}\omega_2^{3^{m-1}}
+\sum_{i=1}^m2^i{m\choose i}\omega_0^{3^{m-1}+\Psi_k(i,m)-1}\omega_1^{3^{m-1}}\omega_2^{3^{m-1}-\Psi_k(i,m)},
\end{eqnarray*}
where $\Psi_k(i,m)$ is the Lloyd polynomial.
\end{thm}
\[\text{\small Table III: the weight distribution of $C_{\overline{g}_{(m,k)}}$ in Theorem \ref{thm2}}\]
 $$
\scalebox{0.9}{

\begin{tabular}{|l|l|l|}\hline
Weight $w$ & Multiplicity $A_{w}$ &condition  \\\hline
0 &1 &$u=0,\mathbf{v}=\mathbf{0}$\\\cline{1-3}
$3^m-3^{m-1}-\Psi_k(i,m)$ &$2^{1+i}{m\choose i}(1\leq i\leq m)$&$u\in\mathbb{F}_3^\ast,wt(\mathbf{v})=i$ with $Re(\widehat{\overline{g}}_{(m,k)}(\mathbf{v}))=\frac{3}{2}(\Psi_k(i,m))$\\\cline{1-3}
 $3^m-\sum_{j=0}^k2^j{m\choose j}$&$2$&$u\in\mathbb{F}_3^\ast,\mathbf{v}=\mathbf{0}$ with $Re(\widehat{\overline{g}}_{(m,k)}(\mathbf{0}))=-\frac{3^m}{2}+\frac{3}{2}\sum_{j=0}^k2^j{m\choose j}$\\\cline{1-3}
  $3^m-3^{m-1}$&$3^m-1$&$u=0,\mathbf{v}\in{\mathbb{F}_3^m}^\ast$\\\hline
\end{tabular}}\\
$$

$\mathbf{Proof.}$ Note that $\overline{S}(m,k)={\mathbb{F}_3^m}^\ast\setminus S(m,k)$, if $\mathbf{v}=\mathbf{0}$, then by Lemmas \ref{lem3}-\ref{lem6}, we have
\begin{eqnarray*}
\widehat{\overline{g}}_{(m,k)}(\mathbf{0})&=&(3^m-1)\zeta_3+3^m+1-3^m-(\zeta_3-1)\sum_{j=1}^k2^j{m\choose j}\\
&=&(3^m-1)\zeta_3-(\zeta_3-1)\sum_{j=1}^k2^j{m\choose j}+1.
\end{eqnarray*}
Hence,
\begin{eqnarray*}
Re(\widehat{\overline{g}}_{(m,k)}(\mathbf{0}))&=&-\frac{1}{2}(3^m-1)+\frac{3}{2}\sum_{j=1}^k2^j{m\choose j}+1\\
&=&-\frac{3^m}{2}+\frac{3}{2}\sum_{j=1}^k2^j{m\choose j}+\frac{3}{2}.
\end{eqnarray*}
If $wt(\mathbf{v})=i(1\leq i\leq m)$, then by Lemmas \ref{lem3}-\ref{lem6}, we have
\begin{eqnarray*}
\widehat{\overline{g}}_{(m,k)}(\mathbf{v})=1-\zeta_3-(\zeta_3-1)(\Psi_k(i,m)-1)=(1-\zeta_3)\Psi_k(i,m).
\end{eqnarray*}
And so
\[Re(\widehat{\overline{g}}_{(m,k)}(\mathbf{v}))=\frac{3}{2}\Psi_k(i,m),\]
thus the weight distribution of $C_{\overline{g}_{(m,k)}}$ follows from Lemma \ref{lem5}.

The calculation for the complete weight enumerator of $C_{\overline{g}_{(m,k)}}$ is similar to that of $C_{f_{(m,k)}}$, we omit it here.

This completes the proof for Theorem \ref{thm2}.  \hfill$\Box$\\

\begin{rem}
Note that Corollary 3.4 in \cite{Mesnager} presents a formula for the weight values of codewords in $p$-ary code $\mathcal{C}_{f_{\overline{D}}}$, where $f_{\overline{D}}$ is the characteristic function of $\overline{D}$, however, the formula is not specific, and the weight distributions of $\mathcal{C}_{f_{\overline{D}}}$ have not been given. In our Theorem \ref{thm2}, a class of ternary minimal codes is presented. Furthermore, both the weight distributions and the complete weight enumerators of $C_{\overline{g}_{(m,k)}}$ are given, which implies our result is more explicit than Corollary 3.4 in \cite{Mesnager}.
\end{rem}

We need the following lemmas to obtain the parameters of $C_{\overline{g}_{(m,k)}}$ in Theorem \ref{thm2}.
\begin{lemma}\label{lem1}
Let $m$ and $t=\lfloor\frac{m-1}{2}\rfloor$ be integers with $m\geq 16$, then
\[{{m+1}\choose {t}}>3{{m-1}\choose {t}}.\]
\end{lemma}
$\mathbf{Proof.}$ Note that
\begin{eqnarray}\label{shizi1}
{{m+1}\choose {t}}-3{{m-1}\choose {t}}=\frac{(m-1)(m-2)\cdots(m-t)}{t!}\left[\frac{(m+1)m}{(m-t+1)(m-t)}-3\right],
\end{eqnarray}
so we can consider (\ref{shizi1}) in two cases.

{\bf Case 1}. If $m=2m_1+1$, then $t=\lfloor\frac{m-1}{2}\rfloor=m_1$ with $m_1\geq 8$, and so we have
\begin{eqnarray*}
\frac{(m+1)m}{(m-t+1)(m-t)}-3&=&\frac{(2m_1+2)(2m_1+1)}{(m_1+2)(m_1+1)}-3\\
&=&\frac{4m_1+2}{m_1+2}-3=4-\frac{6}{m_1+2}-3\\
&=&1-\frac{6}{m_1+2}>0.
\end{eqnarray*}

{\bf Case 2}. If $m=2m_1$, then $t=\lfloor\frac{m-1}{2}\rfloor=m_1-1$ with $m_1\geq 8$, and so we have
\begin{eqnarray*}
\frac{(m+1)m}{(m-t+1)(m-t)}-3&=&\frac{(2m_1+1)2m_1}{(m_1+2)(m_1+1)}-3\\
&=&1-\frac{10m_1+8}{(m_1+2)(m_1+1)}=1-\frac{2}{m_1+2}\frac{5m_1+4}{m_1+1}\\
&=&1-\frac{2}{m_1+2}(5-\frac{1}{m_1+1})>1-\frac{10}{m_1+2}\geq 0.
\end{eqnarray*}

Thus, by {\bf Cases 1-2}, we complete the proof for Lemma \ref{lem1}.\hfill$\Box$\\

\begin{lemma}\label{lem2}
Let $m,k$ be integers with $m\geq 5 $ and $1\leq k\leq \lfloor\frac{m-1}{2}\rfloor $, then
\begin{eqnarray}\label{shizi2}
3^{m-1}-2^k{{m-1}\choose k}-\sum_{j=0}^k2^j{m\choose j}>0.
\end{eqnarray}
\end{lemma}
$\mathbf{Proof.}$ Note that the value of the formula
\[3^{m-1}-2^k{{m-1}\choose k}-\sum_{j=0}^k2^j{m\choose j}\]
decreases with $k$ increasing, and so for $t=\lfloor\frac{m-1}{2}\rfloor$, we have
\begin{eqnarray*}
&&3^{m-1}-2^k{{m-1}\choose k}-\sum_{j=0}^k2^j{m\choose j}\\
&\geq&(1+2)^{m-1}-2^t{{m-1}\choose t}-\sum_{j=0}^t2^j{m\choose j}\\
&=&\sum_{j=0}^{m-1}2^j{{m-1}\choose j}-2^t{{m-1}\choose t}-\sum_{j=1}^t2^j[{{m-1}\choose j}+{{m-1}\choose {j-1}}]-1\\
&=&\sum_{j=1}^{m-1}2^j{{m-1}\choose j}-\sum_{j=1}^t2^j{{m-1}\choose j}-\sum_{j=0}^{t-1}2^{j+1}{{m-1}\choose j}-2^t{{m-1}\choose t}\\
&=&\sum_{j=t+1}^{m-1}2^j{{m-1}\choose j}-\sum_{j=0}^{t-1}2^{j+1}{{m-1}\choose j}-2^t{{m-1}\choose t}\\
&=&\sum_{j=0}^{m-t-2}2^{m-1-j}{{m-1}\choose j}-\sum_{j=0}^{t-1}2^{j+1}{{m-1}\choose j}-2^t{{m-1}\choose t}\\
&\geq&\sum_{j=0}^{t-1}2^{j+1}(2^{m-2j-2}-1){{m-1}\choose j}-2^t{{m-1}\choose t}\\
&>&\sum_{j=0}^{t-1}2^{j+1}{{m-1}\choose j}-2^t{{m-1}\choose t}\\
&>&2^{t-1}{{m-1}\choose {t-2}}+2^{t}{{m-1}\choose {t-1}}-2^{t-1}{{m-1}\choose {t-2}}\\
&=&2^{t-1}[{{m+1}\choose {t}}-3{{m-1}\choose {t}}].
\end{eqnarray*}
By Lemma \ref{lem1}, the inequality (\ref{shizi2}) holds when $m\geq 16$, and for $5\leq m\leq 15$,  one can easily verify (\ref{shizi2}) holds, here we omit the details.

This completes the proof for Lemma \ref{lem2}.\hfill$\Box$\\
\begin{lemma}\label{lem7}
Let $m$ be an integer with $m\geq 2$, then
\[\sum_{j=\lfloor\frac{m-1}{2}\rfloor+1}^{m-1}2^j{{m-1}\choose j}-\sum_{j=0}^{\lfloor\frac{m-1}{2}\rfloor-1}2^{j+1}{{m-1}\choose j}>0.\]
\end{lemma}
$\mathbf{Proof.}$ We prove Lemma \ref{lem7} in two cases.

If $m=2m_1(m_1\geq 1)$ is even, then we have
\begin{eqnarray*}
&&\sum_{j=\lfloor\frac{m-1}{2}\rfloor+1}^{m-1}2^j{{m-1}\choose j}-\sum_{j=0}^{\lfloor\frac{m-1}{2}\rfloor-1}2^{j+1}{{m-1}\choose j}\\
&=&\sum_{m_1}^{2m_1-1}2^j{{2m_1-1}\choose j}-\sum_{j=0}^{m_1-2}2^{j+1}{{2m_1-1}\choose j}\\
&=&\sum_{j=0}^{m_1-1}2^{2m_1-1-j}{{2m_1-1}\choose {2m_1-1-j}}-\sum_{j=0}^{m_1-2}2^{j+1}{{2m_1-1}\choose j}\\
&=&\sum_{j=0}^{m_1-1}2^{2m_1-1-j}{{2m_1-1}\choose j}-\sum_{j=0}^{m_1-2}2^{j+1}{{2m_1-1}\choose j}\\
&>&\sum_{j=0}^{m_1-2}{{2m_1-1}\choose j}(2^{2m_1-1-j}-2^{j+1})>0.
\end{eqnarray*}

If $m=2m_1+1(m_1\geq 1)$ is odd, then we have
\begin{eqnarray*}
&&\sum_{j=\lfloor\frac{m-1}{2}\rfloor+1}^{m-1}2^j{{m-1}\choose j}-\sum_{j=0}^{\lfloor\frac{m-1}{2}\rfloor-1}2^{j+1}{{m-1}\choose j}\\
&=&\sum_{m_1+1}^{2m_1}2^j{{2m_1}\choose j}-\sum_{j=0}^{m_1-1}2^{j+1}{{2m_1}\choose j}\\
&=&\sum_{j=0}^{m_1-1}2^{2m_1-j}{{2m_1}\choose {2m_1-j}}-\sum_{j=0}^{m_1-1}2^{j+1}{{2m_1}\choose j}\\
&=&\sum_{j=0}^{m_1-1}2^{2m_1-j}{{2m_1}\choose j}-\sum_{j=0}^{m_1-1}2^{j+1}{{2m_1}\choose j}\\
&>&\sum_{j=0}^{m_1-1}{{2m_1}\choose j}(2^{2m_1-j}-2^{j+1})>0.
\end{eqnarray*}

By the discussions above, we complete the proof for this Lemma.\hfill$\Box$\\

Now, we give the parameters of $C_{\overline{g}_{(m,k)}}$ in Theorem \ref{thm2} as follows.
\begin{cor}\label{cor1}
Let $m,k$ be integers with $m\geq 5$ and $2\leq k\leq \lfloor\frac{m-1}{2}\rfloor$, then the linear code $C_{\overline{g}_{(m,k)}}$ in Theorem \ref{thm2} has parameters
\[\left[3^m-1,m+1,3^m-3^{m-1}-2^k{{m-1}\choose k}\right].\]
Furthermore, $\frac{w_{min}}{w_{max}}\leq \frac{2}{3}$ if and only if
\[2\sum_{j=1}^k2^j{m\choose j}\leq 3\cdot 2^k{{m-1}\choose k}.\]
\end{cor}
$\mathbf{Proof.}$ By Table III, the weight value for nonzero codewords in $C_{\overline{g}_{(m,k)}}$ has the following three cases,
\begin{eqnarray*}
\left\{\begin{array}{lll}w(i)=3^m-3^{m-1}-\Psi_k(i,m); \\
 w'=3^m-\sum_{j=0}^k2^j{m\choose j};\\
 w''=3^m-3^{m-1}. \end{array}\right.
\end{eqnarray*}
Thus we have
\begin{eqnarray*}
w'-w''&=&3^m-\sum_{j=0}^k2^j{m\choose j}-3^m+3^{m-1}=3^{m-1}-\sum_{j=0}^k2^j{m\choose j}\\
&=&\sum_{j=0}^{m-1}2^j{{m-1}\choose j}-\sum_{j=0}^k2^j{m\choose j}=\sum_{j=k+1}^{m-1}2^j{{m-1}\choose j}+\sum_{j=0}^k2^j{{m-1}\choose j}-\sum_{j=0}^k2^j{m\choose j}\\
&=&\sum_{j=k+1}^{m-1}2^j{{m-1}\choose j}-\sum_{j=1}^k2^j{{m-1}\choose {j-1}}=\sum_{j=k+1}^{m-1}2^j{{m-1}\choose j}-\sum_{j=0}^{k-1}2^{j+1}{{m-1}\choose j}.
\end{eqnarray*}
Note that the value of
\[\sum_{j=k+1}^{m-1}2^j{{m-1}\choose j}-\sum_{j=0}^{k-1}2^{j+1}{{m-1}\choose j}\]
decreases with $k$ increasing, thus by Lemma \ref{lem7} we have
\begin{eqnarray*}
w'-w''&\geq&\sum_{j=\lfloor\frac{m-1}{2}\rfloor+1}^{m-1}2^j{{m-1}\choose j}-\sum_{j=0}^{\lfloor\frac{m-1}{2}\rfloor-1}2^{j+1}{{m-1}\choose j}>0.
\end{eqnarray*}
Note that
\begin{eqnarray*}
w'-w(i)&=&3^m-\sum_{j=0}^k2^j{m\choose j}-3^m+3^{m-1}+\Psi_k(i,m)=3^{m-1}+\Psi_k(i,m)-\sum_{j=0}^k2^j{m\choose j}.
\end{eqnarray*}
Thus according to Lemma \ref{lem4}, one can get
\begin{eqnarray*}
3^{m-1}+\Psi_k(i,m)-\sum_{j=0}^k2^j{m\choose j}\geq&3^{m-1}-2^k{{m-1}\choose k}-\sum_{j=0}^k2^j{m\choose j}.
\end{eqnarray*}
Hence by Lemma \ref{lem2}, we have
\begin{eqnarray*}
w'-w(i)=3^{m-1}+\Psi_k(i,m)-\sum_{j=0}^k2^j{m\choose j}\geq3^{m-1}-2^k{{m-1}\choose k}-\sum_{j=0}^k2^j{m\choose j}>0.
\end{eqnarray*}
From the discussions above, the maximum Hamming weight of $C_{\overline{g}_{(m,k)}}$ is given by $w_{max}=w'$. According to Lemma \ref{lem4}, the minimum Hamming weight of $C_{\overline{g}_{(m,k)}}$ is given by
\[w_{min}=w(1)=3^m-3^{m-1}-\Psi_k(1,m)=3^m-3^{m-1}-2^k{{m-1}\choose k}.\]

This completes the proof for Corollary \ref{cor1}. \hfill$\Box$\\
\begin{rem}
The minimal distance of $C_{g_{(m,k)}}$ is $\sum_{j=1}^k2^j{m\choose j}$, and the minimal distance of $C_{\overline{g}_{(m,k)}}$ is $3^m-3^{m-1}-2^k{{m-1}\choose k}$, so by Lemma \ref{lem2} we have
\[3^m-3^{m-1}-2^k{{m-1}\choose k}-\sum_{j=1}^k2^j{m\choose j}>3^{m-1}-2^k{{m-1}\choose k}-\sum_{j=0}^k2^j{m\choose j}>0.\]
This means that the minimal distance of $C_{\overline{g}_{(m,k)}}$ is better than that of codes constructed in \cite{Heng-Ding-Zhou}.
\end{rem}
The following result shows that $C_{\overline{g}_{(m,k)}}$ in Theorem \ref{thm2} is minimal.
\begin{thm}\label{thm3}
Let $m,k$ be integers with $m\geq 5$ and $2\leq k\leq \lfloor\frac{m-1}{2}\rfloor$. Then $C_{\overline{g}_{(m,k)}}$ is a minimal code with parameters
\[\left[3^m-1,m+1,3^m-3^{m-1}-2^k{{m-1}\choose k}\right].\]
Furthermore, $\frac{w_{min}}{w_{max}}\leq \frac{2}{3}$ if and only if
\[2\sum_{j=1}^k2^j{m\choose j}\leq 3\cdot 2^k{{m-1}\choose k}.\]
\end{thm}
$\mathbf{Proof.}$ According to Corollary \ref{cor1}, we only need to prove that $C_{\overline{g}_{(m,k)}}$ is minimal. From the proof for Theorem \ref{thm2}, we have
\begin{eqnarray}\label{shizi5}
Re(\widehat{\overline{g}}_{(m,k)}(\mathbf{v}))=\left\{\begin{array}{ll}-\frac{3^m}{2}+\frac{3}{2}\sum_{j=0}^k2^j{m\choose j}, &\text{if}\quad \mathbf{v}=\mathbf{0};\\
 \frac{3}{2}\Psi_k(i,m), &wt(\mathbf{v})=i>0. \end{array}\right.
\end{eqnarray}
On the other hand, Lemma \ref{lem8} implies that $C_{\overline{g}_{(m,k)}}$ is minimal if and only if both
\begin{eqnarray}\label{shizi3}
Re(\widehat{\overline{g}}_{(m,k)}(\mathbf{v_1}))+Re(\widehat{\overline{g}}_{(m,k)}(\mathbf{v_2}))-2Re(\widehat{\overline{g}}_{(m,k)}(\mathbf{v_3}))\neq 3^m
\end{eqnarray}
and
\begin{eqnarray}\label{shizi4}
Re(\widehat{\overline{g}}_{(m,k)}(\mathbf{v_1}))+Re(\widehat{\overline{g}}_{(m,k)}(\mathbf{v_2}))+Re(\widehat{\overline{g}}_{(m,k)}(\mathbf{v_3}))\neq 3^m
\end{eqnarray}
hold for any pairwise distinct vectors $\mathbf{v_1},\mathbf{v_2},\mathbf{v_3}\in \mathbb{F}_3^m$ with $\mathbf{v_1}+\mathbf{v_2}+\mathbf{v_3}=\mathbf{0}$. 

We distinguish the following two cases to show that both (\ref{shizi3}) and (\ref{shizi4}) hold for the claimed vectors.

{\bf Case 1}. Only one of $\mathbf{v_1},\mathbf{v_2},\mathbf{v_3}$ is $\mathbf{0}$.

We first consider the inequality (\ref{shizi4}). Without loss of the generality, we assume that $\mathbf{v_1}=\mathbf{0}$ and then $\mathbf{v_2}=-\mathbf{v_3}\neq \mathbf{0}$, where $wt(\mathbf{v_2})=wt(\mathbf{v_3})=i$ ($1\leq i\leq m$). Then by (\ref{shizi5}), we know that (\ref{shizi4}) is equivalent to
\begin{eqnarray*}
-\frac{3^m}{2}+\frac{3}{2}\sum_{j=0}^k2^j{m\choose j}+3\Psi_k(i,m)\neq 3^m\Leftrightarrow\frac{3}{2}\sum_{j=0}^k2^j{m\choose j}+3\Psi_k(i,m)\neq\frac{3}{2}\cdot 3^m.
\end{eqnarray*}
Note that $\sum_{j=0}^k2^j{m\choose j}<3^{m-1}$ for $2\leq k\leq \lfloor\frac{m-1}{2}\rfloor$, then $\frac{3}{2}\sum_{j=0}^k2^j{m\choose j}<\frac{3^m}{2}$. Thus by Lemma \ref{lem4} we have
\begin{eqnarray*}
\frac{3}{2}\sum_{j=0}^k2^j{m\choose j}+3\Psi_k(i,m)&<&\frac{3^m}{2}+3\Psi_k(i,m)<\frac{3^m}{2}+3\cdot 2^k{{m-1}\choose k}\\
&<&\frac{3^m}{2}+3\cdot \sum_{j=0}^{m-1}2^j{{m-1}\choose j}=\frac{3}{2}\cdot 3^m.
\end{eqnarray*}
Next we prove (\ref{shizi3}) in two cases.

(I.1)\quad If $\mathbf{v_3}=\mathbf{0}$, then $wt(\mathbf{v_1})=wt(\mathbf{v_2})=i$ ($1\leq i\leq m$). Thus by (\ref{shizi5}), (\ref{shizi3}) is equivalent to
\begin{eqnarray*}
3\Psi_k(i,m)+3^m-3\sum_{j=0}^k2^j{m\choose j}\neq 3^m\Leftrightarrow \Psi_k(i,m)\neq \sum_{j=0}^k2^j{m\choose j}.
\end{eqnarray*}
Thus by Lemma \ref{lem4},
\[\Psi_k(i,m)< 2^k{{m-1}\choose k}<2^k{m\choose k}<\sum_{j=0}^k2^j{m\choose j},\]
namely (\ref{shizi3}) holds.

(I.2)\quad If one of $\mathbf{v_1}$ and $\mathbf{v_2}$ is $\mathbf{0}$, assume that $\mathbf{v_1}=\mathbf{0}$ without loss of the generality. Then $wt(\mathbf{v_2})=wt(\mathbf{v_3})=i$ ($1\leq i\leq m$). Thus by (\ref{shizi5}), we know that (\ref{shizi3}) is equivalent to
\begin{eqnarray*}
-\frac{3^m}{2}+\frac{3}{2}\sum_{j=0}^k2^j{m\choose j}+\frac{3}{2}\Psi_k(i,m)-3\Psi_k(i,m)\neq 3^m,
\end{eqnarray*}
which holds if and only if
\begin{eqnarray*}
\sum_{j=0}^k2^j{m\choose j}-\Psi_k(i,m)\neq 3^m.
\end{eqnarray*}
Note that
\begin{eqnarray*}
|\sum_{j=0}^k2^j{m\choose j}-\Psi_k(i,m)|&\leq& \sum_{j=0}^k2^j{m\choose j}+|\Psi_k(i,m)|\leq\sum_{j=0}^k2^j{m\choose j}+ 2^k{{m-1}\choose k}\\
&<&\sum_{j=0}^k2^j{m\choose j}+ 2^k{{m-1}\choose k}+2^k{{m-1}\choose {k+1}}=\sum_{j=0}^k2^j{m\choose j}+ 2^k{{m}\choose {k+1}}\\
&<&\sum_{j=0}^k2^j{m\choose j}+ 2^{k+1}{{m}\choose {k+1}}<\sum_{j=0}^m2^j{m\choose j}=3^m.
\end{eqnarray*}
Thus (\ref{shizi3}) holds.

In this case, both (\ref{shizi3}) and (\ref{shizi4}) follow from the discussions above.

{\bf Case 2}. None of $\mathbf{v_1},\mathbf{v_2},\mathbf{v_3}$ is nonzero.

Due to Lemma \ref{lem4} and (\ref{shizi5}), we derive that
\[|Re(\widehat{\overline{g}}_{(m,k)}(\mathbf{v_1}))+Re(\widehat{\overline{g}}_{(m,k)}(\mathbf{v_2}))-2Re(\widehat{\overline{g}}_{(m,k)}(\mathbf{v_3}))|\leq 6\times 2^k{{m-1}\choose k}, \]
and
\[|Re(\widehat{\overline{g}}_{(m,k)}(\mathbf{v_1}))+Re(\widehat{\overline{g}}_{(m,k)}(\mathbf{v_2}))+Re(\widehat{\overline{g}}_{(m,k)}(\mathbf{v_3}))|\leq \frac{9}{2}\times 2^k{{m-1}\choose k}.\]
To show that both (\ref{shizi3}) and (\ref{shizi4}), it is sufficient to show that
\[2^{k+1}{{m-1}\choose k}<3^{m-1}.\]
In fact, by $2\leq k\leq \lfloor\frac{m-1}{2}\rfloor$, we have
\begin{eqnarray*}
2^{k+1}{{m-1}\choose k}&=&2^k{{m-1}\choose k}+2^k{{m-1}\choose k}\\
&<&2^k{{m-1}\choose k}+2^{k+1}{{m-1}\choose {k+1}}\\
&<&\sum_{j=0}^{m-1}2^j{{m-1}\choose j}=3^{m-1}.
\end{eqnarray*}
Thus both (\ref{shizi3}) and (\ref{shizi4}) hold in this case.

Summarizing the discussions above, we complete the proof for Theorem \ref{thm3}. \hfill$\Box$\\

Based on Magma's program, the following example is presented, which is accordant with Theorem \ref{thm2}.
\begin{exmp}
Let $m=9$ and $k=2$, then the code $C_{\overline{g}_{(m,k)}}$ in Theorem \ref{thm2} is a minimal code with parameters $[19682,10,13010]$ and the weight enumerator
\[1+36z^{13010}+288z^{13052}+1344z^{13085}+1024z^{13094}+4032z^{13109}+4608z^{13115}\]
\[+19682z^{13122}+8064z^{13124}+10752z^{13130}+9216z^{13133}+2z^{19520}.\]
Thus $\frac{w_{min}}{w_{max}}=\frac{13010}{19520}<\frac{2}{3}$.
\end{exmp}

\section{Conclusions}
$ $

For an odd prime $p$, using the generic construction (\ref{shizi6}) to obtain minimal codes violating the Ashikhmin-Barg condition, $f$ is usually taken to be the characteristic function of a subset in $\mathbb{F}_p^m$. There are a few known results when $f$ is a non-characteristic function. In this paper, basing on the generic construction (\ref{shizi6}), taking $f$ be the function defined on special sets of vectors in $\mathbb{F}_3^m$, we present two new classes of minimal ternary linear codes violating the Ashikhmin-Barg condition and then determine their complete weight enumerators. Especially, the minimal distance of $C_{\overline{g}_{(m,k)}}$ in Theorem \ref{thm2} is better than that of the code constructed in \cite{Heng-Ding-Zhou}.

 \end{document}